\documentclass[apj,numberedappendix]{emulateapj}
\usepackage{epsfig}
\usepackage{amsmath}
\usepackage{enumerate}
\usepackage{natbib}
\usepackage{hyperref}
\usepackage{ulem}
\usepackage[squaren, Gray, cdot]{SIunits}

\usepackage{color}

\def\mprot{m_{\rm p}}
\def\me{m_{\rm e}}
\def\tGeV{T_p}

\def\tobs{t_{\rm obs}}

\def\NGeV{N_{\rm GeV}}
\def\M{{\cal M}}
\def\Zpm{Z_{\pm}}
\def\Gjet{\Gamma_{\rm ej}}

\def\EIC{E_{\rm IC}}
\def\EKN{E_{\rm KN}}

\def\Rdec{R_{\rm dec}}

\def\ginj{\gamma_{\rm inj}}

\def\UB{U_{\rm B}}
\def\Urad{U_{\rm rad}}
\def\epsB{\varepsilon_{\rm B}}

\def\betapre{\beta_{\rm pre}}
\def\gpre{\gamma_{\rm pre}}
\def\mue{\mu_{\rm e}}
\def\epse{\varepsilon_{\rm e}}
\def\epsnth{\varepsilon_{\rm nth}}

\def\gopt{\gamma_{\rm opt}}
\def\fsyn{f_{\rm syn}}

 \def\Gbw{\Gamma}

\def\Ekin{{\cal E}_{\rm kin}}
\def\EGRB{{\cal E}_{\rm GRB}}
\def\EMeV{{\cal E}_{\rm MeV}}
\def\tauT{\tau_{\rm T}}

\def\LE{L_E}

\newbox\grsign \setbox\grsign=\hbox{$>$} \newdimen\grdimen \grdimen=\ht\grsign
\newbox\simlessbox \newbox\simgreatbox \newbox\simpropbox
\setbox\simgreatbox=\hbox{\raise.5ex\hbox{$>$}\llap
     {\lower.5ex\hbox{$\sim$}}}\ht1=\grdimen\dp1=0pt
\setbox\simlessbox=\hbox{\raise.5ex\hbox{$<$}\llap
     {\lower.5ex\hbox{$\sim$}}}\ht2=\grdimen\dp2=0pt
\setbox\simpropbox=\hbox{\raise.5ex\hbox{$\propto$}\llap
     {\lower.5ex\hbox{$\sim$}}}\ht2=\grdimen\dp2=0pt
\def\simgt{\mathrel{\copy\simgreatbox}}
\def\simlt{\mathrel{\copy\simlessbox}}

\shortauthors{I. Vurm, R. Hasco\"et, A. M. Beloborodov}

\begin{document}

\title{Pair-dominated GeV-optical flash in GRB 130427A}

\author{Indrek Vurm\altaffilmark{1,2}, Romain Hasco\"et\altaffilmark{1} and Andrei M. Beloborodov\altaffilmark{1}}
\affil{$^1$Physics Department and Columbia Astrophysics Laboratory, Columbia University, 538 West 120th Street, New York, NY 10027, USA; indrek.vurm@gmail.com \\
$^2$Tartu Observatory, T\~{o}ravere 61602, Tartumaa, Estonia\\
}

\label{firstpage}
\begin{abstract}
We show that the light curve of the double GeV+optical flash in GRB~130427A 
is consistent with radiation from the blast wave in a wind-type medium
with density parameter $A=\rho r^2\sim 5\times 10^{10}$~g~cm$^{-1}$.
The peak of the flash is emitted by copious $e^\pm$ pairs created and 
heated in the blast wave; 
our first-principle calculation determines
the pair-loading factor and temperature of the shocked plasma. 
Using detailed radiative transfer simulations we reconstruct the observed
double flash.
The optical flash is dominated by synchrotron emission from the thermal
plasma behind the forward shock, and the GeV flash is produced via 
inverse Compton (IC) scattering by the same plasma.
The seed photons for IC scattering are dominated by the prompt MeV radiation 
during the first tens of seconds, and by the optical to X-ray afterglow thereafter.
IC cooling of the thermal plasma behind the forward 
shock reproduces all GeV data from  a few seconds to $\sim 1$~day. 
We find that the blast wave Lorentz factor at the peak of the flash is 
$\Gamma\approx 200$, and the forward shock magnetization is 
$\epsB\sim 2\times 10^{-4}$. 
An additional source is required by the data 
in the optical and X-ray bands at times $>10^2$~s; we speculate that this 
additional source may be a long-lived reverse shock in the explosion ejecta.
\end{abstract}



\section{Introduction}

GRB~130427A was
an exceptionally bright gamma-ray burst due to its relative proximity
(cosmological redshift $z=0.34$, \citealt{Levan13})
and high luminosity reaching
$L_{\rm MeV}\sim 3\times 10^{53}~\mbox{erg s}^{-1}$ in the MeV band 
(\citealt{Ackermann14}, hereafter A14;
\citealt{Golenetskii13}).
The burst was accompanied by a GeV flash with peak luminosity
$L_{\rm GeV} \sim 10^{51}~\mbox{erg s}^{-1}$ (A14)
and an optical flash with peak luminosity $L_{\rm O}\sim 10^{49}~\mbox{erg s}^{-1}$
\citep{Vestrand14}.
It is the first gamma-ray burst (GRB) observed at early times $\tobs<100$~s
by both optical and GeV telescopes. 

Remarkably, the optical and GeV flashes peaked at 
approximately the same time $\tobs\sim 15$~s, and
both showed a smooth decay after the peak;
the optical flux decay $F_\nu\propto t^{-1.67}$ 
was steeper than that in the GeV band. 
Such double (optical+GeV) flashes were predicted  
to result from copious $e^\pm$ pair creation in the blast wave of
the GRB explosion (\citealt{BHV13}, hereafter B13).
In this Letter, we apply this model to GRB~130427A.

In our model, the GeV emission is produced by inverse Compton (IC) cooling 
of the blast wave \citep[see also][]{B05a,Fan05}.
The observed spectrum extends
to at least $\sim 100$~GeV, with a 95~GeV 
photon detected at $243$~s and a $32~\mbox{GeV}$ photon at 34 ks.
Such high-energy photons cannot be produced by synchrotron emission
(\citealt{deJager92,PiranNakar10}; A14; \citealt{Fan13}), which makes a strong 
case for their IC origin.

We calculate the synchrotron and IC cooling of the 
plasma heated in the forward shock of the explosion
using the Monte Carlo radiative transfer code developed in B13.
The code self-consistently solves the coupled problem of radiative transfer,
pair creation, and blast wave dynamics.
The original version of the code included only the prompt radiation as a source
of target photons for IC scatterings; here we also include the optical
to X-ray afterglow radiation, which dominates seed photons for IC scattering at late times.
The prompt and afterglow radiation densities 
used in our calculations are taken from observations.


\section{GeV flash}

\subsection{Pair-dominated peak}

The external medium ahead of the blast wave is 
exposed to the prompt GRB
radiation, which pre-accelerates the medium and loads it with copious
$e^\pm$ pairs \citep{ThompsonMadau00,B02a}.
Bright bursts $e^\pm$-enrich the external medium by a factor $Z_\pm\gg 1$
at radii $R < 10^{17}$~cm.
B13 showed that this effect leads to a bright GeV+optical flash.
The forward shock heats the pair-enriched
medium to the thermal Lorentz factor  given by
\begin{align}
       \ginj \approx \frac{\Gbw}{\gpre(1+\betapre)}\left( 1 + \epse\frac{\mue\mprot}{\Zpm\me} \right),
\label{eq:ginj}
\end{align}
where $\Gbw$ is the blast wave Lorentz factor,
$\gpre$ is the pre-acceleration Lorentz factor
of the $e^\pm$-loaded medium ahead of the blast wave,
$\betapre = (1-1/\gpre^2)^{1/2}$,
$\mue$ is the ion mass per proton in units of $\mprot$
($\mue=1$ for hydrogen and $2$ for heavier elements), and
$\epse$ is the fraction of shocked ion energy transferred to leptons;
B13 showed that at early times $\epse\approx 1$.
In our numerical model presented below we assume $\epse=1$ as long as 
$\Zpm>500$; at later times we take $\epse=0.3$, as suggested by plasma shock 
simulations \citep{SironiSpitkovsky2011}.

The pair-loading factor $Z_\pm$ steeply decreases at $R\simgt 10^{16}$~cm
and hence $\ginj$ grows (Figure \ref{fig:dyn}). This implies a steep rise in the energy
of the IC photons, $\EIC\sim \ginj^2 E_t$, where $E_t$ are the energies
of the seed/target photons. As long as the blast wave overlaps with the prompt
radiation, the seed radiation is dominated by the prompt photons with 
$E_t\simlt 1$~MeV. 
The onset (and peak) of the GeV flash marks the moment when
$\EIC$ reaches the GeV band.
This occurs when $\ginj$ exceeds $\sim 30$.

The condition $\ginj\sim 30$ together with the observed peak time $T_p$
determines the radius and Lorentz factor of the blast wave (B13).
For GRB~130427A we find,
\begin{align}
      R_p= 1.6\times 10^{16} \, \left( \frac{\EGRB}{8\times 10^{53} \, {\rm erg}}\right)^{1/2}
     {\rm cm},
\end{align}

\begin{align}
   \Gbw(R_p) \approx 150 \left( \frac{\EGRB}{8\times 10^{53} \, {\rm erg}}\right)^{1/4}
   \left( \frac{\tGeV}{15 \, {\rm s}}\right)^{-1/2}.
\label{eq:gamma}
\end{align}
Here $\EGRB$ is normalized to the energy of the main prompt MeV episode \citep{Golenetskii13}
and we have used $z=0.34$ \citep{Levan13}.

Assuming that the external medium is a wind from the massive progenitor of the
burst, the expected number of GeV photons in the peak of the flash is (B13),
\begin{align}
    \NGeV \sim 8\times 10^{51} \, \Zpm\,\frac{A_{11} R_{16}}{\mu_e} \, \M,
\label{eq:Nem}
\end{align}
where
$A = \rho R^2=\mbox{const}$ is the wind density parameter,
and $\M\sim 5-10$ is the multiplicity of photons emitted
above $100$ MeV by a single fast-cooling electron.
The pair-loading factor $\Zpm$ steeply drops
from $10^3$ to $10^2$ at
$R\approx R_p$  (Figure \ref{fig:dyn}).
Comparing Equation~(\ref{eq:Nem}) with the observed $\NGeV \sim 5\times 10^{54}$
(A14, \citealt{Fan13}), we conclude that $A=10^{10}-10^{11}$~g~cm$^{-1}$ is required.
Our detailed transfer simulations show that $A\sim 5\times 10^{10}$~g~cm$^{-1}$ gives a
GeV flash that is close to the observed one.

The simulated GeV light curve is shown in the upper panel of Figure \ref{fig:LC};
the corresponding high-energy spectra at 
five time intervals are plotted in Figure \ref{fig:spec}.
The emission above $100$~MeV is initially soft, but quickly hardens as $\ginj$ 
exceeds 30 and then the spectrum remains roughly flat in $\nu F_{\nu}$.
The maximal photon energy,
$E_{\rm IC,max} = \me c^2 \Gamma\ginj (1+z)^{-1}$,
evolves to the TeV range within a few dynamical times as $\Zpm$ drops.

\subsection{Blast wave deceleration}

\begin{figure}[h]
\begin{center}
\includegraphics[trim = 0.5cm 1.5cm 0.5cm 1cm, width=0.45\textwidth]{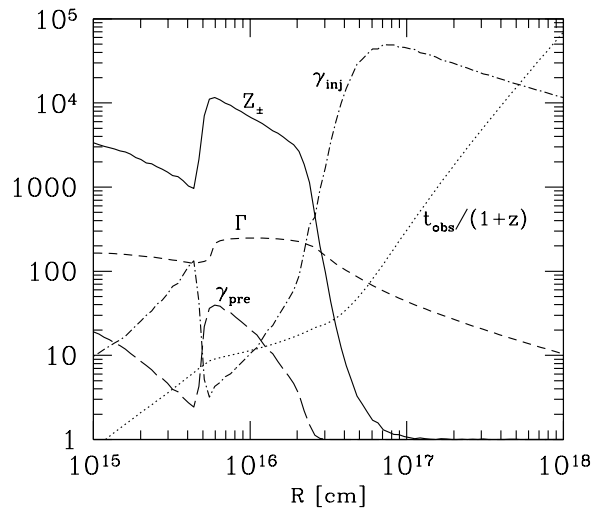}
\caption{Various quantities at the forward shock as a function of radius:
pair loading factor $\Zpm$ (solid line),
pre-acceleration Lorentz factor $\gpre$ (long-dashed line),
blast wave Lorentz factor $\Gbw$ (short-dashed line), 
and electron injection Lorentz factor $\ginj$ (dash-dotted line).
The shapes of $Z_\pm(R)$ and $\gpre(R)$ are controlled by the details of the 
observed prompt emission and obtained numerically, as explained in B13.
Dotted curve shows $\tobs/(1+z)$ (in seconds), where $\tobs$ is
the arrival time of photons emitted at angle $\theta=\Gamma^{-1}$.
}
\label{fig:dyn}
\end{center}
\end{figure}

\begin{figure}[h]
\begin{center}
\includegraphics[trim = 0.5cm 1.5cm 0.5cm 0cm, width=0.45\textwidth, height=0.5\textwidth]{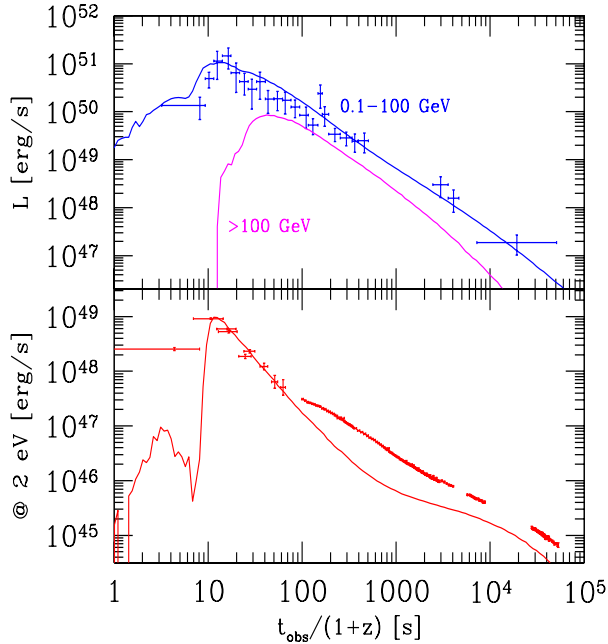}
\caption{The simulated light curves and data for the GeV 
(upper panel) and optical (lower panel) flashes in GRB 130427A.
Parameters:
wind density $A=5 \times 10^{10} \mbox{g cm}^{-1}$, ejecta kinetic energy $\Ekin = 2.5\times 10^{53}$~erg,
forward shock magnetization $\epsB= 2 \times 10^{-4}$,
Lorentz factor of the unshocked jet $\Gjet=350$.
The GeV data is from A14,
the optical data is from \citet{Vestrand14} and \citet{Perley13}.}
\label{fig:LC}
\end{center}
\end{figure}

\begin{figure}[h]
\begin{center}
\includegraphics[trim = 0.5cm 1.5cm 0.5cm 1cm, width=0.45\textwidth]{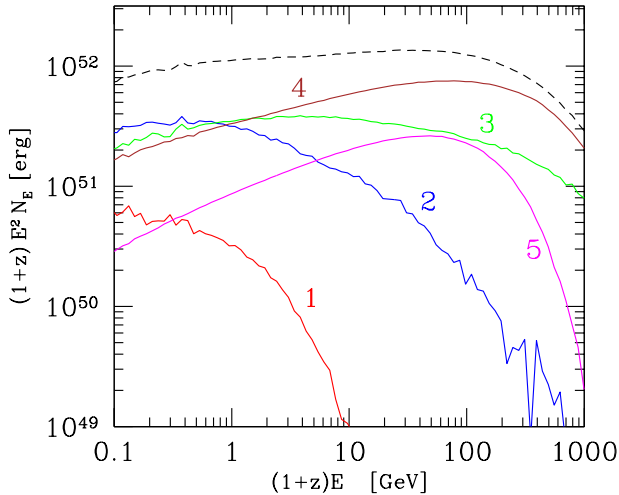}
\caption{
High-energy radiation spectrum
in five time intervals: 
$0-10$~s (1, red), $10-30$~s (2, blue), $30-100$~s (3, green),
   $100-3000$~s (4, brown), and $3000-80000$~s (5, magenta).
The black dashed line shows the total fluence spectrum.
}
\label{fig:spec}
\end{center}
\end{figure}

Our model for the GeV flash gives the parameters $A$, $R_p$, and 
$\Gamma_p$, and implies the explosion energy
\begin{equation}
  \Ekin \approx 2.5\times 10^{53} \,\mbox{erg}.
\end{equation}
It is consistent with a high
radiative efficiency of the prompt emission,
$\EMeV/(\Ekin+\EMeV)\approx 0.8$.
If the prompt emission is considered as a proxy for the ejecta power,
one infers that most of the ejecta kinetic energy is contained
in a shell of material about 15 light-seconds thick.
The (formal) deceleration radius of the blast wave is
\begin{align}
  \Rdec &= 5 \times 10^{15} \, \frac{{\cal E}_{\rm kin}}{2.5\times 10^{53}\, {\rm erg}} \nonumber \\
  &\times \left(\frac{A}{5\times 10^{10} \, {\rm g}\, {\rm cm}^{-1}}\right)^{-1} \left(\frac{\Gamma}{200}\right)^{-2} \, {\rm cm}.
\end{align}
The corresponding timescale $\Rdec/(2c\Gamma^2)$ is shorter than the duration of the prompt emission. Therefore
the reverse shock in this explosion must be relativistic;
it crosses the shell in approximately the same time 
as it takes the main prompt episode to completely overtake the forward shock.
The reverse shock crossing marks the time when the bulk of the jet kinetic energy has been
transferred to the blast wave.
At this point the blast wave is still radiatively efficient 
(as the pair loading factor $\Zpm$ is still high);
the explosion loses a substantial fraction of its initial 
energy during the first $20$-$100$~s. 
This results in the steep decline of $\Gbw(R)$ 
at $R\sim (2-5)\times 10^{16}$~cm (Figure \ref{fig:dyn}).
At $t\sim 100$~s the blast wave approaches the adiabatic self-similar regime
with $\Gamma \propto [\Ekin/(A t)]^{1/4}$ and $R \propto (\Ekin t/A)^{1/2}$.


\subsection{Transition to synchrotron-self-Compton cooling}
 \label{sec:SSC}

The prompt radiation decouples from the forward shock at $\sim 30$~s and
does not contribute to IC cooling at later times. Then the blast wave is mainly
cooled by IC scattering of the afterglow radiation, which is produced by the blast
wave itself via synchrotron emission \citep[see also][]{Liu13,Tam13}.
Remarkably, the transition to the ``synchrotron-self-Compton'' (SSC) phase
is smooth, with no easily recognizable feature in the GeV light curve (Figure \ref{fig:LC}).
The main reason for this is that the electrons at this stage are still in the fast-cooling regime,
which renders their IC emission insensitive to the target photon luminosity.

At the beginning of the SSC phase, $\ginj$ is already high and
the IC scattering is dominated by low-energy photons,
below the Klein-Nishina energy,
\begin{align}
    \EKN
   \approx \frac{\Gamma\,  \me c^2}{\ginj (1+z)}  =  
   1\, \varepsilon_{{\rm e}, -1}^{-1} \, {\rm keV},
\label{eq:E_KN}
\end{align}
where we have used Equation (\ref{eq:ginj}) with $\Zpm=1$ and $\gpre=1$,
as pair creation is weak at late times.
Equation (\ref{eq:E_KN}) along with $\EIC \propto E_t$ implies that the
energies of target photons upscattered to the LAT band
range from optical to soft X-rays.
We approximate the spectral luminosity of the target (afterglow) radiation as 
\begin{align} 
   L_E=L_E^0\,
   \left(\frac{E_t}{1 \,\mbox{keV}}\right)^{-\alpha} t_3^{-\beta},
   \label{eq:AG}
\end{align}
where $t=\tobs/(1+z)$.
We use  $L_E^0=3\times 10^{56}~\mbox{s}^{-1}$, 
$\alpha = 0.55$ for the optical to X-ray spectral 
index, and $\beta=1.1$ for the temporal index \citep[e.g.][]{Perley13}.

The IC cooling time of the thermal plasma behind the forward shock 
becomes longer than the dynamical time at $\sim 10^4$~s.
Our numerical calculations show that the transition
to the slow-cooling regime is very gradual with no easily identifiable spectral or
temporal signature in the GeV emission (Figure \ref{fig:LC}).

The decay slope of the high-energy light curve
cannot be described by a simple analytical model.
Naively, in the fast-cooling stage one would expect
$\EIC\LE(\EIC) \propto \epse L_{\rm diss} ( \EIC/E_{\rm IC,max})^{-\alpha+1}$,
where $L_{\rm diss}=8\pi c^3 A\Gamma^4$ is the total luminosity dissipated at the shock,
yielding $\EIC \LE (\EIC)
\propto \varepsilon_{\rm e}^{0.55} \Ekin^{0.78} A^{0.22} \EIC^{0.45} t^{-0.78}$.
Similarly, in the slow-cooling phase
$\EIC\LE(\EIC) \propto \tauT \ginj^2 E_t 
L_E(E_t) \propto
\varepsilon_{\rm e}^{1.1} \Ekin^{-0.23} A^{1.2} \EIC^{0.45} t^{-1.9}$.
The simulated light curve
is inconsistent with either regime and decays approximately as $t^{-1.2}$ up to $\sim 10^4~\mbox{s}$.

This behavior results from a few effects. In the fast-cooling phase the temporal decay is steeper
than the naive prediction due to the contribution from secondary pairs
produced by the partial absorption of the GeV flash; this effect
declines with time and becomes negligible at a few $100~\mbox{s}$.
The decreasing pair loading (up to $\sim 100$~s) also somewhat steepens the 
light-curve. Furthermore, the large-angle GeV radiation from the main peak affects 
the observed light curve after the peak.
The gradual transition to the slow cooling regime around $\sim 10^4~\mbox{s}$
results in a broad bump in the light curve as electrons start accumulating 
at $\ginj$; the asymptotic slow cooling regime is only approached at $t\gtrsim 1$~d.

The maximum energy of IC photons produced by the thermal electron population,
$E_{\rm IC,max} = 270 \, \varepsilon_{{\rm e},-1} {\cal E}_{{\rm kin},54}^{1/2}  A_{11}^{-1/2}  t_3^{-1/2} \mbox{GeV}$,
can accommodate the observed multi-GeV photons at late times,
in particular the $32~\mbox{GeV}$ photon observed at 34 ks.

\subsection{TeV emission}

The relative proximity and high luminosity of GRB 130427A makes it
an interesting target for very high energy (VHE) observations.
Our model predicts emission of photons of energies $\sim 1$~TeV.
The simulated VHE light curve above 100 GeV is shown in the top panel of 
Figure~\ref{fig:LC} (magenta line).
The luminosity above $100$~GeV 
reaches the peak of
$\sim 8\times 10^{49}~\mbox{erg s}^{-1}$
during the first minute, and most of the
VHE fluence should be received in $\sim 1000$~s.
Such flashes are detectable with current Cerenkov telescopes.
To our knowledge no rapid VHE follow-up was performed 
for GRB~130427A by presently operating observatories.
VERITAS obtained an upper limit at $\sim 1$~d, which indicates 
a (temporal or spectral) break when compared with the extrapolation of the 
earlier LAT observation below 100~GeV (J. McEnery, private communication).
This is consistent with our model, as the
predicted VHE emission from the thermal electrons behind the shock
cuts off at about $50$~ks, when the characteristic
IC photon energy falls below 100~GeV.

\section{Optical flash}

The optical flash is produced by synchrotron emission from the same thermal 
electrons injected at the forward shock that give rise to the GeV emission (B13).
The mechanism of the delayed onset, peak and early decay is also
analogous. The bright optical flash occurs when the  synchrotron emission reaches
the optical band as the electron injection Lorentz factor increases,
i.e. when $\ginj=\gopt$, where
\begin{align}
\gopt = \left( \frac{5 E}{\Gamma h\nu_{B}} \right)^{1/2}  
	&\approx 500
	\, (\varepsilon_{{\rm B}, -3} A_{11})^{-1/4} \, R_{16}^{1/2} \, \Gamma_{2}^{-1} \nonumber \\
	&\times\left[ \gpre(1+\betapre) \right]^{1/4} \, \left( \frac{E}{2 \, {\rm eV}} \right)^{1/2},
\end{align}
and $\nu_{B}=eB/(2\pi \me c)$ is the cyclotron frequency.

The energetic pairs behind the shock are in the fast-cooling regime at the 
peak of the flash.
As the electron/positron cools, most of the optical radiation is emitted
when its Lorentz factor  $\gamma\sim \gopt$.
The approximate optical luminosity is given by (B13)
\begin{align}
L_{\rm O}\approx   \Zpm \frac{dN_{\rm p}}{dR} \frac{dR}{dt} \frac{\me c^2 \gopt \Gamma}{2}  \fsyn,
\label{eq:Lopt}
\end{align}
where the factor
$\fsyn \approx \UB/\Urad$ accounts for the fraction of energy radiated as synchrotron emission.
The observed optical luminosity near the peak, $\sim 10^{49}$ erg/s,
requires $\epsB\sim 10^{-4}-10^{-3}$.

The theoretical optical light curve at $2~\mbox{eV}$ is plotted in the 
lower panel of Figure \ref{fig:LC}.
Compared to the GeV flash, the onset is slightly delayed,
because the threshold $\ginj$ for producing 
synchrotron optical radiation is somewhat higher than that for  
producing IC GeV radiation.
The decay of the optical
flash is controlled by the declining pair loading factor
$\Zpm$ and is consistent with the observed light curve
up to $\sim 100$~s.
At later times synchrotron emission from nonthermal electrons 
must take over, which is not included in the model shown in Figure~2.


\section{Discussion}

The
observed GeV flash in GRB~130427A can be
explained as IC emission from the thermal plasma behind the blast wave in a wind
medium, once the pair loading of the blast wave is correctly taken into account.
The same model reproduced the GeV flash in GRB~080916C (B13).
The exceptional LAT data for GRB~130427A, which extends to $\sim 1$~d,
made it possible to test the model at longer times, when the seed photons for IC
scattering change from the prompt radiation to the afterglow. We found that this
transition leaves no sharp features and is consistent with the entire observed light
curve of GeV emission.

The hot $e^\pm$ plasma in the blast wave must also emit synchrotron radiation,
in particular in the optical band. The predicted optical light curve is very close to the
optical flash observed during the first 100~s (Figure~\ref{fig:LC}).
This provides further support to the proposed model.

Figure~2, the main result of this paper, shows only
emission from the {\it thermal} plasma behind the forward shock, which is a robust 
consequence of shock heating and is straightforward to model from first principles.
We also performed a simulation including
a nonthermal population of leptons in the forward shock, with an injection spectrum $dN_{\rm inj}/d\gamma\propto \gamma^{-p}$ (with $p=2.2$)
carrying a fraction $\epsnth=0.1$ of the shock energy.
We found that the additional synchrotron and IC radiation produced by 
this nonthermal component weakly affects the predicted
GeV+optical flash. 
We conclude that the thermal postshock plasma dominates the flash, 
at least in the region of parameter space explored by our simulations.
A higher $\epsnth$ and a flat electron spectrum $p\approx 2$ would make 
the contribution from nonthermal particles more significant, especially 
before the peak of the flash, making the rise toward the peak less sharp. 
Detailed models
with thermal+nonthermal shocked plasma are deferred to a future paper.

After the peak, the synchrotron frequency of the thermal electrons heated by the forward shock
$\nu_{\rm syn,th}$ remains above the optical band until $\sim 10^4$~s.
In this situation, the addition of nonthermal electrons with $\gamma>\ginj$
does not significantly increase the optical emission from the forward shock.
The additional (nonthermal) contribution to the optical afterglow observed at
$10^2$-$10^4$~s should be produced by a different source, most
likely a long-lived reverse shock \citep{Uhm07,Genet07}.
This agrees with the suggestion of 
previous works on GRB~130427A \citep{Panaitescu13,Laskar13,Perley13}.

Our flash model requires the wind density parameter
$A\sim 5\times 10^{10}$~g~cm$^{-1}$.  It is much higher (and more typical of 
Wolf-Rayet stars) than previously inferred from 
nonthermal afterglow modeling at $t>10$~min
\citep{Panaitescu13,Laskar13,Perley13}.
A constraint on $A$ from the late afterglow comes from the following consideration.
When the characteristic synchrotron frequency of the forward shock crosses 
the optical band (which happens at $t\sim 10^4$~s in our model)
its predicted optical flux is
$F_\nu\sim  2 \, A_{11}(\varepsilon_e/0.3)^{-2}(\mu_e/2)^{-3}\, t_4$~mJy.
It should not exceed the observed flux of 2 mJy, 
a condition satisfied by our model.
Models assuming $\mu_e=1$ (hydrogen) and $\epsilon_e=0.1$ 
require smaller $A$, in agreement with Panaitescu et al. (2013).
One should also keep in mind that the wind density profile may deviate from $R^{-2}$,
i.e. the effective $A$ may change in the late afterglow.
Yet we find no conflict between $A\sim 5\times 10^{10}$~g~cm$^{-1}$ and radio data
at $t>1$~d; the blast wave can produce radio emission without 
significant self-absorption in the forward or reverse shock.

Our model implies a high radiative
efficiency of the blast wave at early times, when pair loading is strong.
During the first few $100$~s the blast wave energy
$\Ekin$ drops from $2.5\times 10^{53}$~erg to $\sim 10^{53}$~erg. 
We expect that in a more detailed model a long-lived reverse shock 
will add energy to the blast wave and keep $\Ekin$ from falling to such low values.
A few lines of evidence suggest this energy injection. 
First, this would help to explain the high X-ray luminosity.
Without additional energy 
the power dissipated in the forward shock is low,
\begin{align}
    L_{\rm diss} = \Ekin/4t = 2.5\times 10^{49} \, E_{{\rm kin}, 53} \, t^{-1}_3 \,\mbox{erg s}^{-1}.
\end{align}
It is only a factor of 9 higher than the observed $0.3$-$10$~keV luminosity at
$t\gtrsim 10^{3}$~s, which would require a very high efficiency of X-ray emission.
Secondly, the observed X-ray spectral index
indicates that the (nonthermal) electrons are radiating 
X-rays in the slow cooling regime already at $\sim 1000$~s.
At these early times, electrons are mainly cooled by IC scattering (not synchrotron) 
and the cooling frequency $\nu_{\rm syn,c}$ is very sensitive to the blast wave energy,
$\nu_{\rm syn,c}\propto \epsB^{1/2}\Ekin^p t^q$,
where $p=(4+\alpha)/2\alpha\approx 4.1$, $q = (4\beta-3\alpha)/2\alpha\approx 2.5$,
and $\alpha$, $\beta$ are the afterglow spectral and temporal indices
defined in Equation~(\ref{eq:AG}).
Energy injection via the reverse shock 
would help to keep 
$\nu_{\rm syn,c}$ above the X-ray band.
We find that supplying $\Ekin\sim 10^{54}$~erg by $t\sim 1000$~s 
may be sufficient to explain the slow-cooling regime in 
the X-ray band. This can be accomplished by a tail of the GRB jet
with $\Gamma_{\rm tail}\approx 50-100$ carrying energy comparable to the jet head.

The increased $\Ekin$ will boost the optical luminosity, which can overshoot
the observed afterglow,
in particular when $\nu_{\rm syn,th}$ crosses the optical band at $t\sim 10^4$~s.
This problem could be resolved
if $\epsB$ is reduced by a factor of $\sim30$ by that time.
It is not unreasonable to assume that $\epsB$ evolves, as physical conditions
change in the expanding blast wave; e.g., the pair loading is quickly decreasing.
Another factor that can reduce $\epsB$ is the increasing cooling length of the 
shock-heated plasma. Note that $\epsB$ describes the 
{\it average} value of
the magnetic field in the emission region and depends on how quickly the field decays
downstream of the shock \citep[e.g.][]{Lemoine13}.

The reduction of $\epsB$ in the late afterglow phase is also suggested by the
high value of the cooling frequency, 
$\nu_{\rm syn,c}$, inferred from observations by
NuSTAR at $t\sim 1$~d. NuSTAR
identified a break at $\sim 100$~keV in the afterglow spectrum,
which was interpreted as a cooling break \citep{Kouveliotou13}.
With no evolution of $\epsB$, our model would predict the break at a few keV while
a reduction of $\epsB$ by a factor of $\sim 10$ between $10^2$ and $10^5$~s would
move the cooling break to $\sim 100$~keV
(note that cooling at 1~d is dominated by synchrotron emission, not by IC scattering,
and therefore $\nu_{\rm syn,c}\propto \epsB^{-3/2} \Ekin^{1/2} t^{1/2}$).

Detailed modeling of the nonthermal optical and X-ray emission from the forward 
and long-lived reverse shocks is
an involved problem, which we defer to a future work.
It should not, however, change the results of the present paper.
We emphasize that both the optical flash and the {\it entire} GeV 
light curve are insensitive to the details of
energy injection and the evolution of $\epsB$.
The same is true for our estimate of the wind density parameter $A$.

\acknowledgements
This work was supported by NSF grant AST-1008334
and NASA Fermi Cycle 6 grant NNX 13AP246.


\end{document}